# Spin-dependent metalens with intensity-adjustable dual-focused vortex beams

Qun Hao, Wenli Wang, Yao Hu[*], Shaohui Zhang, Shuo Zhang, Yu Zhang

*Beijing Key Laboratory for Precision Optoelectronic Measurement Instrument and Technology, School of Optics and Photonics, Beijing Institute of Technology, Beijing 100081, China*

*Corresponding author: huy08@bit.edu.cn

**Abstract**

Vortex beams with orbital angular momentum has been attracting tremendous attention due to their considerable applications ranging from optical tweezers to quantum information processing. Metalens, an ultra-compact and multifunctional device, provide a desired platform for designing vortex beams. A spin-dependent metalens can boost the freedom to further satisfy practical applications. By combining geometric phase and propagation phase, we propose and demonstrate an approach to design a spin-dependent metalens generating dual-focused vortex beams along longitudinal or transverse direction, i.e., metalenses with predesigned spin-dependent phase profiles. Under the illumination of an elliptical polarization incident beam, two spin-dependent focused vortex beams can be observed, and the relative focal intensity of them can be easily adjusted by modulating the ellipticity of the incident beam. Moreover, we also demonstrated that the separate distance between these dual-focused beams and their topological charges could be simultaneously tailored at will, which may have a profound impact on optical trapping and manipulation in photonics.

## 1. Introduction

Vortex beams carrying orbital angular momentum (OAM) have been a research hotspot after their discovery by Allen et al. in 1992 [1]. These OAM-carrying beams contain the helical phase fronts, with an azimuthal angular dependence of $\exp(-im\alpha)$, where $m$ is the topological charge, and $\alpha$ is the azimuthal angle [2]. The topological charge $m$ can take an arbitrary value within a continuous range, resulting in unbounded OAMs [3]. Possessing this feature, vortex beams have considerable application in optical tweezers and spanners [4,5], optical communication [2,6], nonlinear optics [7], and quantum information processing [8,9]. Currently, multiple optical components are designed for vortex beams generation, such as spiral phase plate [10], space light modulators [11], Q-plates [12] and computer-generated holograms [13]. However, these conventional vortex beams generators limit their practical employment because of bulk mass and single-function, especially, when various vortex beams with structured wavefronts are needed. To satisfy the requirements of both miniaturization and multi-function, researchers are encouraged to seek novel vortex beams generators.

Recently, metasurfaces, composed of specifically designed subwavelength units in a two-dimensional plane, have attracted progressively increasing attention due to their unprecedented control over the properties of light [14–17]. Instead of acquiring desired phase changes through propagation effect, it is capable of tailoring space-variant abrupt phase directly [18]. Therefore, a single metasurface has great potential for the combination of functions of several conventional optical components [16,17,19–21], which is suitable to act as miniaturized multifunctional device [22–25]. In these regards, metasurfaces are ideal vortex beams generators [26–28]. Moreover, focused vortex beams have also been demonstrated using metasurfaces [29–33], which are essential ingredients for optical trapping [34] and manipulation [18]. Meanwhiles, highly focused vortex beams with different topological charges are also needed [18].

Up to now, several excellent works on visible- or infrared-wave metasurfaces with two or multiple focused vortex beams carrying different topological charges have been realized [3,18,34–37]. For instance, Teng et al. [35] designed a spatial multiplexing metalens with dual-focused vortex beams. Each vortex beam carries controllable topological charges, and the relative focal intensity of it is fixed. If the relative focal intensity is required to be changed, the metalens need to be repatterned. Chen et al. [36] demonstrated a holographic metasurface with multiple focused vortex beams, whose relative focal intensity are also invariable. In addition, Tian et al. [37] proposed a nonlinear metasurface that can focus the incident beam with various helicities and wavelengths into three diverse vortex beams, simultaneously, and the focal lengths of the three vortex beams are mathematically relevant. For further practical applications, it is highly desired to design a single metasurface, which can simultaneously control multiple focused vortex beams independently (i.e., topological charges and focal lengths or lateral displacements) and supply additional degrees of freedom to control the relative focal intensity of the output vortex beams while maintaining the advantages of planar profiles, compactness, and relative ease of designing.

In this paper, we design a series of spin-dependent metalenses generating dual-focused vortex beams along longitudinal or transverse direction, and propose their corresponding phase equations and design method. To demonstrate the feasibility of our method, we design several metalens cases based on the proposed phase equations, and perform some numerical simulations using commercial software -FDTD Solutions. The theoretical analysis of the phase equations and design method of our metalenses are both displayed in section 2. The corresponding simulated results in section 3 indicate that our metalenses can simultaneously focus different helicity incident beams into different focused vortex beams independently, the separate distance and topological charges of which can be simultaneously designed at will, and the relative focal intensity of which can also be easily adjusted.

## 2. Theory and design of spin-dependent metalenses
### 2.1 *Overall design of the metalenses*

A spin-dependent metalens is designed and schematically shown in Fig.1. The metalens is composed of numerous titanium dioxide nanobricks with diverse rotation orientation angles and diverse cross sizes sitting on a fused silica substrate. Under the illumination of a left-helicity circular polarization (LCP) incident beam, such a metalens can generate a single- focused vortex beam along longitudinal or transverse direction. Note that Fig. 1 only shows a type of transverse distribution of the focused vortex beams. In contrast, the right-helicity circular polarization (RCP) incident beam can be modulated into another focused vortex beam with different topological charges and focal length (along longitudinal distribution) or lateral displacement (along transverse distribution). For linear polarization (LP) or elliptical polarization (EP) incident beam, dual-converged vortex beams can be simultaneously generated by this metalens independently, and the relative focal intensity of which can be easily allocated by controlling the ellipticity of the incident beam (see Fig.1).

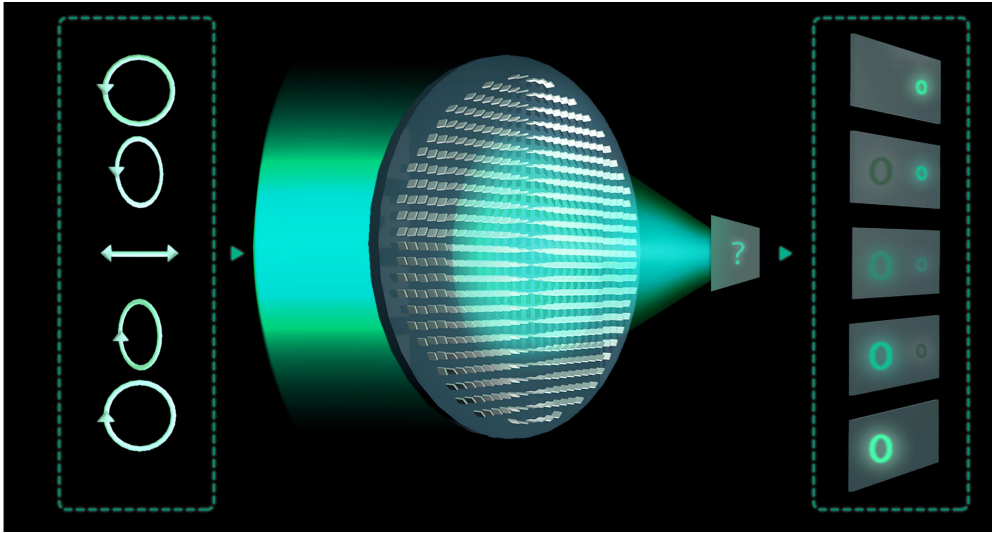

Fig.1 Schematic of the spin-dependent metalens with intensity-adjustable dual-focused vortex beams. The pattern in the left dotted-box represents the polarization state of the monochromatic incident beam; their corresponding monochromatic focused vortex beams are shown in the right dotted-box. Note that the patterns of polarization state and focused vortex beams in these two dotted boxes are in horizontal correspondence.

To design the above-mentioned metalenses, we analyze the phase equations of them and realize the phase with nanobrick structures. The details are proposed as follows.

## 2.2 *Phase equations of the metalenses*

For a metalens that can focus an LCP or RCP incident beam into a single-focused vortex beam, the required phase equation can be respectively calculated via the following expression [38]

$$\varphi(x,y) = -k_0[\sqrt{(x-x_0)^2 + (y-y_0)^2 + f_0^2} - f_0)] + l \cdot \arctan(y/x), \tag{1}$$

where, $x$ and $y$ are the coordinates in the *x-y* plane of the metalens; $k_0 = 2\pi/\lambda$ is the free-space wave vector; $\lambda$ is the wavelength; $f_0$ is the focal length of this metalens, and $l$ is the topological charge. The generated single-converged vortex beam is located at ($x_0$, $y_0$, $f_0$), theoretically, and ($x_0$, $y_0$) is an arbitrary coordinate in the *x-y* plane.

If a metalens can focus an incident beam into two spin-dependent focused vortex beams, the phase profile of this metalens should contain two different phase profiles. Here, inspired by Refs. [24,33], we introduce the helicity of the incident beam into the phase equation, more accurately, into the focal lengths, lateral displacements, and topological charges of the dual-focused vortex beams. Hence, the phase profile of dual-focused vortex- beam metalens appears like this:

$$\varphi_\sigma = -k_0\{\sqrt{[x-(\alpha+\sigma a)]^2 + [y-(\beta+\sigma b)]^2 + (f+\sigma c)^2} - (f+\sigma c)\} \\ +(n+\sigma d)\cdot \arctan(y/x). \qquad (2)$$

Here, $\sigma$ represents the spin of the incident beam. Other variables, i.e., *α, a, β, b, f, c, n,* and *d* are all real numbers. When $\sigma=1$, and then $\varphi_\sigma$ represents the phase profile of the metalens for an LCP incident beam, the corresponding focal length is $f+c$ and the topological charge is $n+d$; the converged vortex beam is located at ($\alpha+a, \beta+b, f+c$), theoretically, and ($\alpha+a, \beta+b$) is a coordinate representing the lateral displacement of this focused vortex beam in the *x-y* plane. Conversely, when $\sigma=-1$, and then $\varphi_\sigma$ represents the phase profile of the metalens for a RCP incident beam, the focal length is $f-c$ and the topological charge is $n-d$; the focused vortex beam is distributed at ($\alpha-a, \beta-b, f-c$) in theory, and ($\alpha-a, \beta-b$) is also a coordinate representing lateral displacement of the focused vortex beam in the *x-y* plane. So, the focal lengths, lateral displacements, and topological charges of dual-focused vortex beams can be simultaneously designed at will.

In particular, when these two converged vortex beams are designed to distributed longitudinally and located at (0, 0) in the *x-y* plane, equation (2) can be simplified as

$$\varphi_\sigma = -k_0[\sqrt{x^2+y^2+(f+\sigma c)^2} - (f+\sigma c)] + (n+\sigma d)\cdot \arctan(y/x), \qquad (3)$$

where, 2*c* represents the separate distance between two focused vortex beams, and 2*d* is the difference of their topological charges.

Similarly, when the metalens is designed to focus the incident beam into transverse dual-focused vortex beams distributed at ($\alpha+\sigma a, 0, f$) or ($0, \beta+\sigma b, f$), respectively, equation (2) can be respectively written as

$$\varphi_\sigma = -k_0\{\sqrt{[x-(\alpha+\sigma a)]^2 + y^2 + f^2} - f\} + (n+\sigma d)\cdot \arctan(y/x), \qquad (4)$$

$$\varphi_\sigma = -k_0\{\sqrt{x^2 + [y-(\beta+\sigma b)]^2 + f^2} - f\} + (n+\sigma d)\cdot \arctan(y/x). \qquad (5)$$

Here, *f*, the neutral distance, becomes the focal length. Equation (4) is the phase profile for the metalens that can focus the incident beam into dual-focused vortex beams along *x*- direction, where 2*a* is the distance between these two converged vortex beams. Equation (5) denotes the phase distribution for the metalens that can focus the incident beam into dual-converged vortex beams along y- direction, where 2b represents the separate distance between the focused beams.

**2.3 Design of the metalenses**

To realize the above phase equations of the metalenses with the practical nanobrick structures, we need to divide the total phase into geometric phase and the propagation phase, which are respectively related with the rotation orientation angles $\theta$ and cross-section sizes (length L, width W) of the nanobricks. If each nanobrick is designed working as a half-wave plate, the incidence of circularly polarized (CP) light can be completely transformed into cross- polarized light. Thus, the rotation orientation angles of geometric phase and propagation phase can be expressed [39] with the phase profiles in equation (2) as:

$$\theta(x,y) = [\varphi_1(x,y) - \varphi_{-1}(x,y)]/4, \tag{6}$$

$$\delta_x(x,y) = [\varphi_1(x,y) + \varphi_{-1}(x,y)]/2, \tag{7}$$

$$\delta_y(x,y) = [\varphi_1(x,y) + \varphi_{-1}(x,y)]/2 - \pi, \tag{8}$$

where $\delta_x$ and $\delta_y$ are propagation phase distributions for an LP incident beam in x- and y-directions, respectively; $\varphi_1$ and $\varphi_{-1}$ represent independent phase profiles for $\sigma = \pm 1$. Then, we pick proper nanobricks whose phase are the most similar with the desired phase to fill in the corresponding positions. This method is also suitable for achieving the phase distributions in equations (3)~(5).

Based on the theory analysis mentioned above, the specifically designed unit cell is shown in Fig.2(a) and (b), which consists of a titanium dioxide nanobrick sitting on a fused silica substrate. These nanobricks are periodically arranged with a fixed square lattice constant $P_x = P_y = 360\text{nm}$, and a height $H = 600\text{nm}$. The geometric phase can easily cover 0~2$\pi$ if the rotation orientation angles $\theta$ of the nanobricks can range from 0 to $\pi$. In addition, each nanobrick is designed to work as a half-wave plate for maximizing the polarization conversion efficiency. Differently, the propagation phase can be modulated by changing the length L and width W of the nanobrick. To make it cover 0~2π, the range of L and W of the nanobricks covers 80 to 280nm, with 2 nm increments of each geometric variable. And their simulated phase shifts $\delta_x$ for x polarized (XLP) incident beam using commercial software FDTD Solutions is shown in Fig.2(c), and the corresponding transmittance distribution is exhibited in Fig.2(d). Note that the phases $\delta_y$ and transmittance distributions for y polarized incident beam can be obtained through rotation of Figs.2(c) and 2(d), respectively [40].

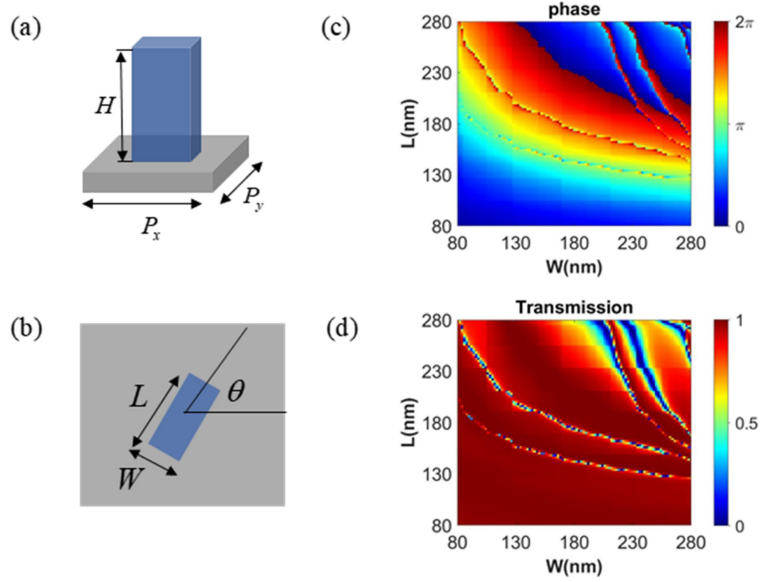

Fig.2 (a) and (b) are side view and vertical view of typical unit cell of the spin-dependent metalens with the period ($P_x$ and $P_y$), height ($H$), varying cross sizes ($L$ and $W$), and different rotation angles ($\theta$). (c) and (d) are simulated propagation phase and transmission distributions for an XLP incident beam, respectively.

## 3. Results and discussions

To demonstrate the proposed phase equations and design method, we design three typical metalenses, whose phase profiles are respectively calculated via equations (3)~(5). We set the polarization of the incident beam as CP, and show the variations in the topological charge and focal length or lateral displacement of the generated single-converged vortex beam when the helicity of the incident beam is changed. Then, we pick one of these designed metalenses to exhibit the change of the relative focal intensity of the dual-focused vortex beams when the ellipticity of the incident beam varies. In addition, according to equations (3)~(5), we design another set of metalenses and set the polarization of the incident beam as XLP. We compare their generated dual-converged vortex beams with those generated by the previous-designed metalenses to demonstrate that topological charges and focal lengths or lateral displacements can be simultaneously tailored arbitrarily.

### 3.1 *Metalens for single-focused vortex beam with alternative topological charges and focal lengths or lateral displacements*

Firstly, we design a metalens (Metalens 1) with a longitudinal single-focused vortex beam, whose focal length and topological charge can be modulated by changing the helicity of the incident beam. According to the equation (3), when the designed metalens is illuminated by LCP, the metalens turns into a single-converged vortex beam metalens with a focal length $f+c$ and topological charges $n+d$. Moreover, the transmitted beam converts to the cross-polarization. i.e., RCP, because the nanobricks are all working as half-waveplates. Similarly, when the polarization of the incident beam becomes RCP, the focal length of this metalens changes to $f-c$; the topological charges switch to $n-d$; the polarization of the beam in the transmission field also transforms into LCP, synchronously. Here, we set $f=10.5$ μm, $c=1.5$ μm, $n=1.5$ and $d=-0.5$, so we could obtain the theoretical parameters of this metalens, which are shown in Table 1. Moreover, this designed metalens consists of $123\times123$ nanobricks covering an area of 490.2 square microns, and the working wavelength of the incident laser is 0.532μm.

The numerical simulation results of Metalens 1 are shown in Fig.3, where Fig.3(a) and (b) represent the intensity patterns in the $x-z$ cross section. The simulated focal lengths are 12.3μm and 9.3μm for an LCP and RCP incident beam, respectively. In our proposed phase profile, we only consider the linear factor of the helicity-dependent focal length. If we take the non-liner factor into account, the simulated focal lengths would be more accurate [24]. The simulated results are basically consistent with the theoretical values in Table 1. In addition, Fig.3(c) and (d) present the simulated converged vortex beams of this metalens in the case of topological charges of 1 and 2. We can conclude that the above metalens can independently manipulate the LCP and RCP incident beams into a single-focused vortex beam with various topological charges.

Table 1. The theoretical parameters of the designed metalenses

|  | Metalens 1 | | Metalens 2 | | Metalens 3 | |
| --- | --- | --- | --- | --- | --- | --- |
| **Incident beam (LCP/RCP)** | LCP | RCP | LCP | RCP | LCP | RCP |
| **Focal lengths (μm)** | 12 | 9 | 9 | 9 | 9 | 9 |
| **Lateral displacements along *x*- direction (μm)** | 0 | 0 | −1 | 0.5 | 0 | 0 |
| **Lateral displacements along *y*- direction (μm)** | 0 | 0 | 0 | 0 | −1 | 0.5 |
| **Topological charges** | 1 | 2 | 1 | 2 | 1 | 2 |

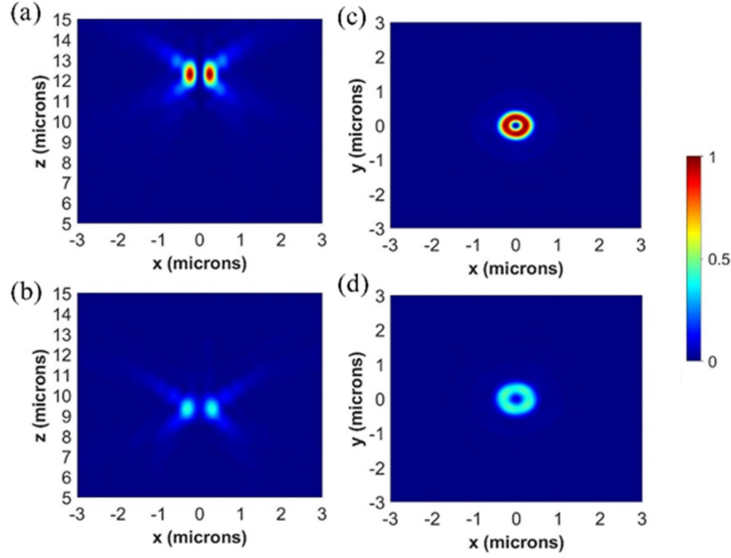

Fig.3 (a) and (b) are simulated electrical field intensity patterns in the $x-z$ plane generated by Metalens1 under illumination of an LCP and RCP incident beam, respectively. (c) and (d) are simulated intensity distributions of the converged vortex beams in different focal planes (z = 12.3μm or 9.3μm) for an LCP and RCP incident beam, respectively.

Similarly, we also design other two metalenses (Metalens 2 and Metalens 3) that could focus LCP/RCP incident beams into RCP/LCP vortex beams with diverse topological charges in the transverse direction, the phase profiles of which are calculated via equation (4) and (5), respectively. Here, the designed parameters are $\alpha = \beta = -0.25$ μm, $a = b = -0.75$ μm, $n = 1.5$, $d = -0.5$ and $f = 9$ μm, and then we could obtain the theoretical parameters of these metalens, which are both shown in Table 1. The simulated results of the Metalens 2 are shown in Fig.4(a) and (b), which indicate two converged vortex beams in the case of topological charges of 1 and 2 are respectively generated under the illumination of LCP and RCP incident beams. They are located at (-1,0,9.3) μm, (0.5,0,9.3) μm, respectively, i.e., exhibiting the designed $x$-direction displacements. The simulated results of the Metalens 3 are shown in Fig.4(c) and (d), which represent the intensity patterns of two focused vortex beams for an LCP and RCP incident beam, respectively. The topological charges of these two vortex beams are 1 and 2, respectively. The positions of them are (0, -1,9.3) μm, (0,0.5,9.3) μm, respectively. i.e., exhibiting the designed $y$-direction displacements. We can see that our designed metalenses could separately focus LCP/RCP incident beams into transverse focused vortex beams with different topological charges.

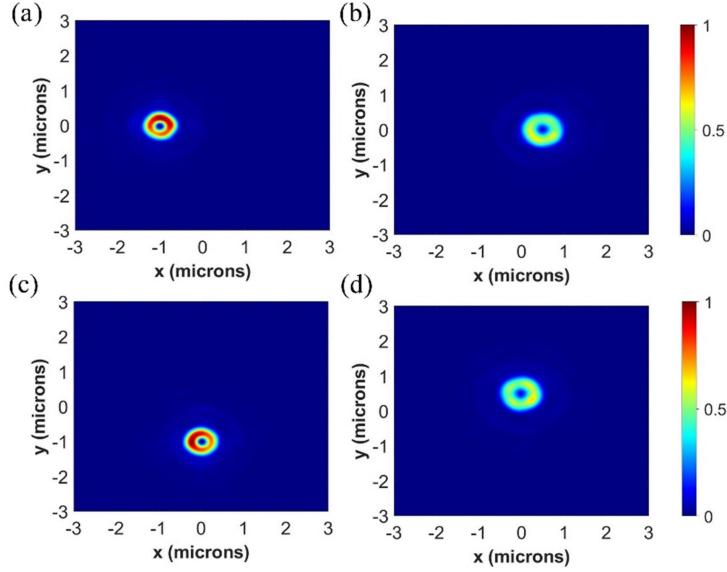

Fig.4 (a) and (b) are simulated intensity distributions of the focused vortex beams generated by Metalens 2 with different *x*-direction displacements in the same focal plane (z = 9.3μm) for LCP and RCP incident beams, respectively. (c) and (d) are simulated intensity distributions of the focused vortex beams generated by Metalens 3 with different *y*-direction displacements in the same focal plane (z = 9.3μm) for LCP and RCP incident beams, respectively.

### 3.2 *Metalens for dual-focused vortex beams with diverse relative focal intensities*

Above simulations have verified the feasibility that two sets of phase profiles used to generate different focused vortex beams could be imparted to a single metalens. Since an LP or EP incident beam can consist of any two orthogonal polarization states, such as LCP and RCP, we infer that the relative focal intensity of these two focused vortex beams can be modified by adjusting the proportion of LCP and RCP of the incident beam. Here, we take Metalens 1 in section 3.1 as an example, where two different focused vortex beams with the topological charges of 1 and 2 are corresponding to the LCP and RCP portion of the incident beam, respectively. Increasing the RCP portion of the incident beam can gain the intensity of the converged vortex beam with the topological charges of 2. In contrast, more portion of LCP in the incident beam will lead to more intensity of the focused vortex beam with the topological charge of 1. In brief, changing the ellipticity of the incident beam can adjust the relative focal intensity of these two focused vortex beams. We calculate their relative focal intensity distributions in the $x-y$ plane for different ellipticity $\chi$ of the incident beam, which is defined as $\chi = (E_{RCP} - E_{LCP})/(E_{RCP} + E_{LCP})$ [24], where $E_{LCP}$ and $E_{RCP}$ respectively represent the intensity of the incident LCP and RCP beam. For simplicity, we exhibit three representative results shown in Fig.5. From Fig.5(a) to Fig.5(c), the ellipticity $\chi$ is set with respective to 1/5, 0, −1/5, and their corresponding LCP ratio of the incident beam increases gradually. Especially, when $\chi$ is equal to 0, the proportion of LCP and RCP is equal. From these figures, it is clear to see that the relative focal intensity between these two vortex beams generated simultaneously can be flexibly modulated by changing the ellipticity of the incident beam, which can be easily achieved by rotating a quarter-wave plate before the metalens. Note that the local intensity of the converged vortex beams will decrease with increasing of topological charges.

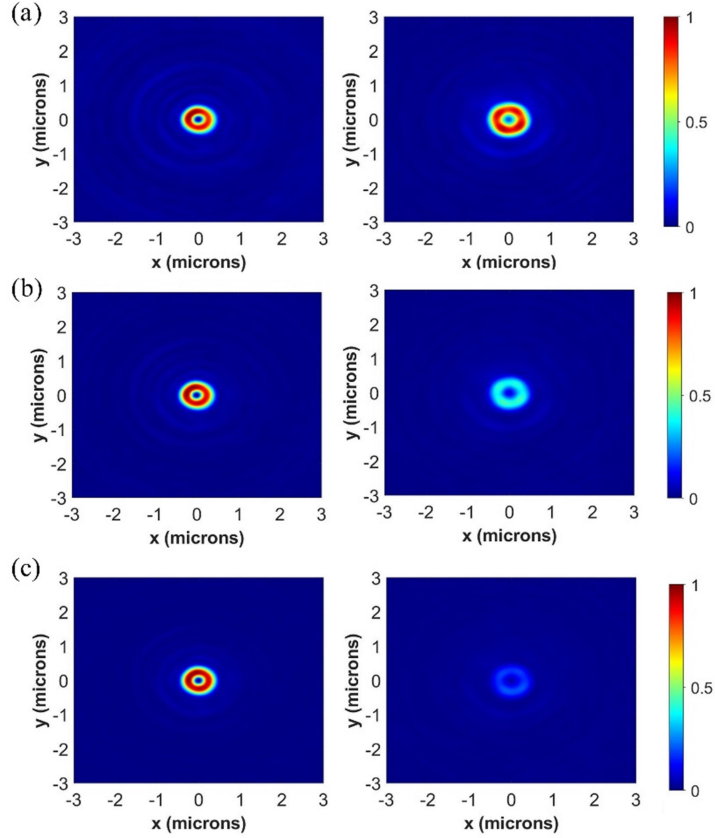

Fig.5 The intensity distributions of the focused vortex beams generated by Metalens 1 in different focal planes (left: z=12.3μm; right: z=9.3μm) when (a) ellipticity of the incident beam is 1/5, (b) ellipticity of the incident beam is 0, and (c) ellipticity of the incident beam is −1/5.

### 3.3 *Metalens for dual-focused vortex beams with different topological charges and focal lengths or lateral displacements*

To verify that the distance between the two focused vortex beams and their topological charges can be simultaneously manipulated at will, we design another three metalenses. Firstly, we take the metalens generating the longitudinal two converged vortex beams as an example. According to our proposed equation in Eq. (3), modifying the value of distance $2c$ between two focused vortex beams and the neutral distance $f$ can tailor arbitrary focal lengths. Similarly, the topological charges can be controlled at will by changing the value of $n$ and $d$. Here we designed another metalens named Metalens 1.1, whose relevant parameters are displayed in Table 2. Comparing Metalens 1 in section 3.1 with Metalens 1.1, the simulated intensity patterns of them in the *x-z* plane are shown in Fig.6(a) and (b) when these two metalenses are both illuminated by XLP incident beam. In this case, the incident beam consists of equal portion of LCP and RCP, so we can obtain two focused vortex beams with almost equal total intensity of bright ring in the focal plane.

In addition, we also take the metalenses with transverse dual-focused vortex beams into consideration. In Eq. (4), if we change the value of the distance $2a$ between two focused vortex beams and the middle position $\alpha$, any two different kinds of lateral displacements can be designed. In Eq. (5), we can also achieve it by controlling the value of $2b$ and $\beta$. The method of manipulating the topological charges of dual-focused vortex beams in the transverse direction is the same as that in the longitudinal direction. Here, the designed parameters of our metalenses (Metalens 2.1 and Metalens 3.1) are displayed in Table 2. Comparing Metalens 2 in section 3.1 with Metalens 2.1, their simulated intensity distributions in the *x-y* plane are respectively displayed in Fig.6(c) and (d) when XLP is incident onto the metalenses. Fig.6(e) and (f) respectively represent the *x-y* plane intensity distributions of Metalens 3 and Metalens 3.1 when illuminated by XLP incident beam. In a word, it is clear to see that the lateral displacements and topological charges of the converged vortex beams between these two sets of metalenses are diverse. The above results confirm that we can independently and arbitrarily control the topological charges and focal lengths or lateral displacements of the two focused vortex beams by adjusting relevant parameters.

Table 2. The theoretical parameters of the designed metalenses

|  | Metalens 1.1 | | Metalens 2.1 | | Metalens 3.1 | |
|---|---|---|---|---|---|---|
| **Incident beam (LCP/RCP)** | LCP | RCP | LCP | RCP | LCP | RCP |
| **Focal lengths (μm)** | 13 | 9 | 9 | 9 | 9 | 9 |
| **Lateral displacements along *x*- direction (μm)** | 0 | 0 | -1.5 | 0.5 | 0 | 0 |
| **Lateral displacements along *y*- direction (μm)** | 0 | 0 | 0 | 0 | -1.5 | 0.5 |
| **Topological charges** | 1 | 3 | 1 | 3 | 1 | 3 |

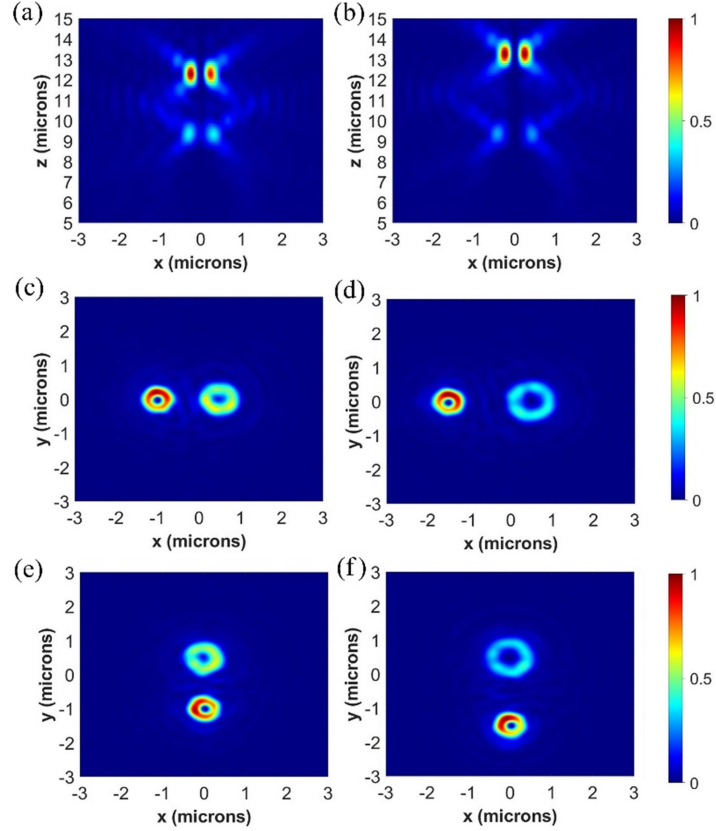

Fig.6 (a) and (b) are the intensity patterns of the longitudinal focused vortex beams respectively generated by Metalens 1 and Metalens 1.1 in the *x-z* plane. (c) and (d) are the intensity patterns of the *x*-direction-distributed focused vortex beams respectively generated by Metalens 2 and Metalens 2.1 in the focal plane. (e) and (f) are the intensity patterns of the *y*-direction-distributed focused vortex beams respectively generated by Metalens 3 and Metalens 3.1 in the focal plane. Note the whole of results are all obtained under the illumination of an XLP incident beam.

## 4. Conclusions

In summary, we have proposed and demonstrated an approach to design a spin-dependent metalens generating dual-focused vortex beams along longitudinal or transverse direction. These spin-dependent metalenses can focus one monochromatic visible wave with different helicity into cross-polarization one or two focused vortex beams. When CP beam is incident on the metalenses, they act as single- converged vortex beam metalenses, whose focal lengths and topological charges or lateral displacements can be changed by changing the helicity of the incident beam. If the incident beam is EP, two different focused vortex beams can be obtained, and the relative focal intensity of these two focused vortex beams can be adjusted by manipulating the ellipticity of the incident beam. Moreover, their separate distance and topological charges can be simultaneously tailored at will. Such designed metalenses may find potential applications in miniaturization and multifunction of optical trapping system and manipulation system.

**Declaration of Competing Interest**

The authors declare that they have no known competing financial interests or personal relationships that could have appeared to influence the work reported in this paper.